\documentclass[amsmath,nobibnotes,aps,superscriptadress,amssymb,pra,aps,showpacs,superscriptaddress,twocolumn]{revtex4-1}
\usepackage{amsmath,amsfonts,amssymb,amsthm,graphics,graphicx,epsfig,bbm}
\usepackage[colorlinks=true,citecolor=blue,linkcolor=blue,urlcolor=blue]{hyperref}
\usepackage[usenames]{color}
\usepackage{graphicx}
\usepackage{subfigure}
\usepackage{amsmath}
\usepackage{epsfig}
\usepackage{dcolumn}
\usepackage{bm}
\usepackage{color}
\usepackage{times}
\usepackage{epstopdf}
\usepackage{amssymb}
\usepackage{amstext}
\usepackage{latexsym}
\usepackage{hyperref}
\usepackage{amsfonts}
\usepackage{psfrag}
\usepackage{soul,xcolor}
\usepackage[normalem]{ulem}
\usepackage{dsfont}
\usepackage{txfonts}
\usepackage{physics}
\usepackage{footnote}
\usepackage{blindtext}
\usepackage{bbm}

\newcommand{\ie}{{\it{i.e.~}}}
\newcommand{\etal}{{\it{et al.}}}

\begin{document}
\setstcolor{red}
	
\title{Digital-analog quantum algorithm for the quantum Fourier transform} 

\author{Ana Martin}
\email[Corresponding authors: ]{\quad mikel.sanz@ehu.es , \quad ana.martinf@ehu.eus}
\affiliation{Department of Physical Chemistry, University of the Basque Country UPV/EHU, Apartado 644, 48080 Bilbao, Spain}

\author{Lucas Lamata}
\affiliation{Department of Physical Chemistry, University of the Basque Country UPV/EHU, Apartado 644, 48080 Bilbao, Spain}
\affiliation{Departamento de F\'isica At\'omica, Molecular y Nuclear, Universidad de Sevilla, 41080 Sevilla, Spain}

\author{Enrique Solano}
\affiliation{Department of Physical Chemistry, University of the Basque Country UPV/EHU, Apartado 644, 48080 Bilbao, Spain}
\affiliation{IKERBASQUE, Basque Foundation for Science, Maria Diaz de Haro 3, 48013 Bilbao, Spain}
\affiliation{International Center of Quantum Artificial Intelligence for Science and Technology (QuArtist) \\ and Department of Physics, Shanghai University, 200444 Shanghai, China}

\author{Mikel Sanz}
\email[Corresponding authors: ]{\quad mikel.sanz@ehu.es , \quad ana.martinf@ehu.eus}
\affiliation{Department of Physical Chemistry, University of the Basque Country UPV/EHU, Apartado 644, 48080 Bilbao, Spain}

\begin{abstract}
Quantum computers will allow calculations beyond existing classical computers. However, current technology is still too noisy and imperfect to construct a universal digital quantum computer with quantum error correction. Inspired by the evolution of classical computation, an alternative paradigm merging the flexibility of digital quantum computation with the robustness of analog quantum simulation has emerged. This universal paradigm is known as digital-analog quantum computing. Here, we introduce an efficient digital-analog quantum algorithm to compute the quantum Fourier transform, a subroutine widely employed in several relevant quantum algorithms. We show that, under reasonable assumptions about noise models, the fidelity of the quantum Fourier transformation improves considerably using this approach when the number of qubits involved grows. This suggests that, in the Noisy Intermediate-Scale Quantum (NISQ) era, hybrid protocols combining digital and analog quantum computing could be a sensible approach to reach useful quantum supremacy.
\end{abstract}
	
\maketitle

\section{Introduction}
Almost four decades ago, a new paradigm, based on laws of quantum mechanics, has been put forward by Y. Manin \cite{Manin1980} and R. Feynman \cite{Feyn1982}. The new paradigm employed quantum features to speed up calculations and it was called quantum simulation or quantum computation (QC). There exist several computational tasks for which QC offers exponential speedups over their classical counterparts \cite{Shor1996,Grover1996}. If we had a fully functional, error corrected quantum computer, we would be able to solve problems that not even the largest classical supercomputers can. But nowadays we are far from this point. The first series of commercial digital quantum processors based on superconducting circuits have been introduced by companies such as IBM, Rigetti, Google and Alibaba. These devices belong to the so-called Noisy Intermediate-Scale Quantum (NISQ) era, in which their performance still faces multiple technical constraints. These constraints pose a great challenge when one tries to solve real-world problems, limiting its size to small-scale \cite{QAlgsBeginers2018, MCRMCLOSS2019}.

Quantum error mitigation (QEM) techniques have been proposed as a possible method to bypass the NISQ-era hardware limitations and improve the calculation of mean values of observables in problems comprising short-depth circuits \cite{TBG17}. These techniques, in general, post-process the information in order to mitigate the effects of the noise \cite{EBL18, CTGSTGC19, KTCMCG19, MYB19}.

A. Parra-Rodriguez \etal introduced in Ref. \cite{PLLSS2018} an alternative hybrid quantum computation approach which could reduce the limitations of NISQ devices. This universal paradigm, called {\it digital-analog quantum computation} (DAQC), merges the flexibility of digital quantum computation with the robustness of analog quantum simulators. If our analog resource is the natural interaction Hamiltonian in the platform, then applying fast single-qubit rotations in certain order, one can generate an arbitrary Hamiltonian. They claim that this codification is susceptible of smaller errors than digital quantum computing when performing quantum simulations. It is noteworthy to mention that, even though the Hamiltonian employed is the Ising model, the DAQC approach is universal with essentially every Hamiltonian \cite{DMNBT2002}.

A natural question is whether quantum algorithms with possible speedup can be efficiently written using this paradigm. The quantum Fourier transform (Q$\mathcal{F}$T) is a key ingredient for several quantum algorithms such as Shor's algorithm for factorization \cite{Shor1996} or the quantum phase estimation algorithm for the estimation of the eigenvalues of a unitary operator \cite{Nielsen2000}. The latter additionally appears as a subroutine of other algorithms, such as the Harrow-Hassidim-Lloyd (HHL) algorithm for linear systems of equations \cite{HHL2009} or the quantum principal component analysis algorithm \cite{LMR2014}. The quantum version of the discrete Fourier transform (D$\mathcal{F}$T) has a exponential speed up over its classical counterpart. While on the classical version it is necessary to apply $\mathcal{O}(n2^n)$ gates, where $n$ refers to the number of bits, on the quantum approach only $\mathcal{O}(n^2)$ gates are needed, in this case $n$ stands for the number of qubits.

In this article, we show how to efficiently write the Q$\mathcal{F}$T algorithm using the DAQC paradigm, and demonstrate that it achieves better results than the purely digital approach on a noisy hardware. For that purpose, we considered the homogeneous all-to-all (ATA) two-body Ising model as a resource for DAQC implementation, and we express the Hamiltonian of the Q$\mathcal{F}$T as an inhomogeneous ATA two-body Ising model. Afterwards, we simulate numerically the cases of a $3-$, $5-$, $6$ and $7-$qubit device, introducing reasonable noise models in the interactions. Additionally, we have performed the Q$\mathcal{F}$T of a certain family of states using both the purely digital and the DAQC approaches. The fidelity between the ideal transformation and the one achieved by the DAQC behaves qualitatively better with the number of qubits than the fidelity offered by the digital implementation. Although this new paradigm has its own noise sources, it eliminates the errors derived from the entangling two-qubit gates. Getting rid of these source of errors allows us to successfully implement relevant quantum algorithms in the NISQ era.

\section{Digital-analog quantum computing}
There are two main approaches to implement QC, namely, the \textit{digital quantum computation} (DQC) and the \textit{analog quantum simulation}. A digital quantum computer, which is based on quantum circuits and the quantum gate model, is a physical platform, such as trapped ions \cite{LHR2011,MMZB2016} or superconducting circuits \cite{BLSM2015,SMLS2015, BSLSNM2016, LSKDBLTG2017, KMTTBChG2017, KDSLS2018}, which can be programmed to efficiently simulate another dynamics of interest. The drawback of this approach is that it consumes too many resources to implement useful applications beyond desired computation, so that it can hardly be considered a viable option with current technology. Some examples of this approach are in the simulation of quantum machine learning \cite{ARSL2018, OSCSL2018}, finance \cite{MCRMCLOSS2019, DLMGLMS2019}, open quantum systems \cite{SSSPS2016}, quantum chemistry \cite{GALMSSL2016}, or quantum field theories \cite{LSKDBLD2017}, among others. On the other hand, analog quantum computing uses a controllable quantum system whose dynamics is known to mimic the dynamics of another system of interest. There are multiple results following this approach simulating, for instance, the quantum Rabi model \cite{BRGDS12, MLHPDSL, PLFRLS2015, BMSSRWU17, LALZPLSK2018}, fluid dynamics \cite{MSLESS2015}, or Casimir physics \cite{FSLRJDS2014, RFERSS2016, SWGS2018}, among others.

Merging these two approaches, leads to a paradigm known as digital-analog quantum computation \cite{GACMELRS2015, APLS2016, LPSS2018, PLLSS2018}. A digital-analog protocol, built combining analog blocks with digital steps, shows the flexibility of the digital gate model \cite{Deutsch1985} and the robustness of the analog simulation model. A formal definition of these elements could be found in Ref.~\cite{PLLSS2018}. Here, we give a practical definition: a digital step is constituted by single-qubit unitary operations whereas an analog block is constituted by the time evolution of a known interaction Hamiltonian.

\begin{figure}
\centering
\includegraphics[width=0.48\textwidth]{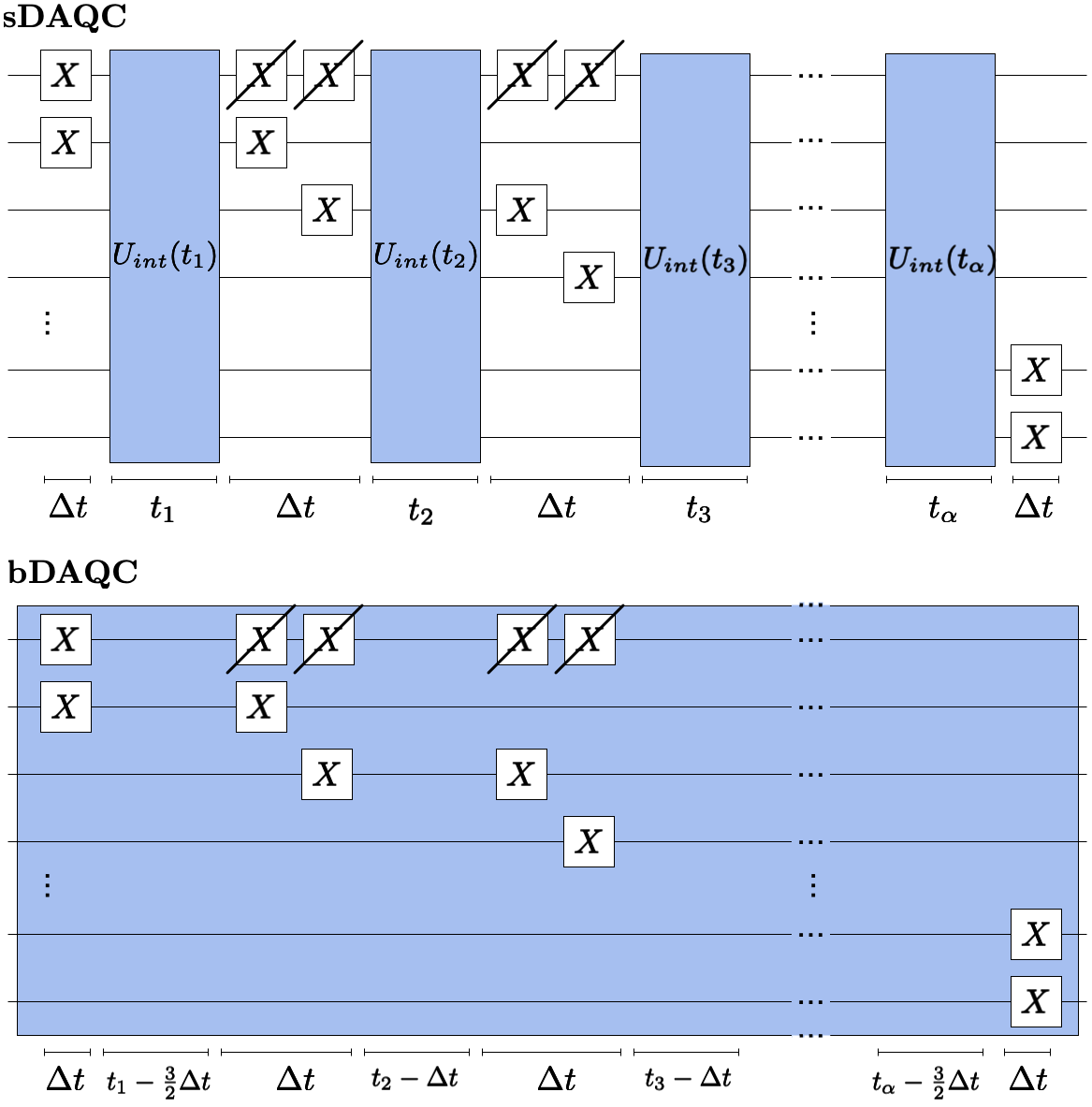} 
\caption{{\bf Comparison between the sDAQC and the bDAQC protocol}. The blue blocks $U_\text{int}(t)$ represent the analog blocks and the single-qubit gates $X$ refers to the Pauli Matrix $\sigma_x$. In the sDAQC protocol, the digital and the analog blocks alternates with each others. The evolution of the interaction Hamiltonian is turned on and off several times. When applying the bDAQC protocol, the analog block is turned on during the whole simulation and the digital blocks are performed on top of the analog evolution.} \label{Fig:sDAQC_vs_bDAQC}
\end{figure}

The most popular quantum processors are based on superconducting circuits where the role of the qubits is played by transmons. The interactions that appears in such physical systems are well described by the inhomogeneous ATA two-body Ising Hamiltonian. Something similar happens with spin-spin interaction in trapped ions. Therefore, from here on, we will use the unitary evolution generated by homogeneous ATA two-body Ising Hamiltonian as the elementary analog block 
\[H_0=H_{\text{int}}=g\sum_{j<k}^N Z^{(j)} Z^{(k)} \rightarrow U_{\text{int}}(t)=e^{i t H_{\text{int}}}, \] 
where $g$ is a fixed coupling strength and $Z^{(i)}$ is the Pauli matrix $\sigma_z^{(i)}$ applied on the $i$-th qubit. For the digital steps, we will employ single-qubit unitary rotations around the X axis with continuous angle $\theta$ between $0$ and $2\pi$ radians. As we will explain below, our goal is to generate an arbitrary ATA inhomogeneous Hamiltonian 
\begin{equation}
H_{ZZ}=\sum_{j<k}^N g_{j k}Z^{(j)} Z^{(k)} \qquad\text{with}\qquad U_{ZZ}=e^{it_F H_{ZZ}}.
\end{equation}
The problem reduces to find an appropriate map between $t_F g_{j,k}$ and $g t_{nm}$ by slicing the homogeneous time evolution $U_{zz}(t)$ into $N(N-1)/2$ analog blocks of different time lengths $t_{nm}$, sandwiched by the local rotations $X^{(n)} X^{(m)}$, as explained in Ref.~\cite{PLLSS2018} and depicted in Fig.~\ref{Fig:sDAQC_vs_bDAQC}. This mapping yields
\begin{eqnarray}
H_{ZZ}&=&\sum_{j<k}^N g_{j,k}Z^{(j)} Z^{(k)}\nonumber\\
&=&\frac{g}{t_F}\sum_{j<k}^N\sum_{n<m}^N t_{nm} X^{(n)} X^{(m)} Z^{(j)} Z^{(k)} X^{(n)} X^{(m)}.
\end{eqnarray}
As $X^{(n)}Z^{(k)}X^{(n)}$ is equal to $-Z^{(n)}$ if $n=k$ and it is equal to $Z^{(k)}$ otherwise, then 
\begin{equation}
H_{ZZ}=\frac{g}{t_F}\sum_{j<k}^N\sum_{n<m}^N t_{nm}(-1)^{\delta_{nj}+\delta_{nk}+\delta_{mj}+\delta_{mk}}Z^{(j)} Z^{(k)}.
\end{equation}

Thus, the problem of finding the value of each time $t_{nm}$ is a matrix-inversion problem
\begin{equation}
g_\beta=t_\alpha M_{\alpha \beta}\frac{g}{t_F}\quad\rightarrow\quad t_\alpha=M_{\alpha\beta}^{-1}g_\beta \frac{t_F}{g},
\end{equation}
where $\alpha$ and $\beta$ are introduced to vectorize each pair of indeces $(n,m)$ and $(j,k)$ as
\begin{equation}
\alpha=N(n-1)-\frac{n(n+1)}{2}+m,\qquad\beta= N(j-1)-\frac{j(j+1)}{2}+k ,
\end{equation}
and $M$ is a sign matrix built up by the elements
\begin{equation}
M_{\alpha\beta}=(-1)^{\delta_{nj}+\delta_{nk}+\delta_{mj}+\delta_{mk}}.
\end{equation}
This sign matrix $M$ is a non-singular matrix $\forall N\in \mathbb{Z}-\{4\}$. This means that, for the case $N=4$ qubits, we need a different set of single qubit rotations. This case is discussed in detail in Ref.~\citep{PLLSS2018}.

The method aforementioned is called \textit{stepwise DAQC} (sDAQC) and, under ideal circumstances, \ie without taking into account noise sources or experimental errors, would lead to the same state as the DQC method. There is another variant of the DAQC method, called \textit{banged DAQC} (bDAQC) protocol. In this case, the analog Hamiltonian is on during the whole simulation and the single-qubit rotations are preformed on top of it. Note that, in the previous case, the analog evolution is turned off before applying single-qubit rotations. The total amount of time in which the analog block is on in the bDAQC, is the sum of the different analog blocks in the sDAQC protocol, as shown in Fig.~\ref{Fig:sDAQC_vs_bDAQC}. 

The bDAQC does not generate the same result as the sDAQC or the DQC method. There is an intrinsic error on the bDAQC which does not depend on either the experimental conditions or noise sources. This error is due to the superposition between the Hamiltonians of the single qubit rotations and the analog Hamiltonian. However, one could expect that, if single qubit rotations are performed in a time $\Delta t$ much smaller than the intrinsic time scale of the analog block, the error will be smaller than the one coming from switching on and off the analog Hamiltonian. Indeed, the additional error per single qubit rotation introduced by not turning off the evolution of the Hamiltonian is of the order $\mathcal{O}\left((\Delta t)^3\right)$ \cite{PLLSS2018}. The reason why we aim at using the bDAQC protocol despite its intrinsic error is because it accumulates less experimental error. Experimentally switching on and off the Hamiltonian is not an exact step function, it takes some time to stabilize. Quantum control tries to suppress these errors, but it turns cumbersome when the system scales up and cannot be solved in a classical computer. If we keep the analog block on during the evolution, we will avoid these errors. This will be of great importance when we explore a more realistic implementation of the DAQC protocol in section \ref{sec:noise}.

\section{Quantum Fourier transform: description and ideal case implementation}
D$\mathcal{F}$T plays an important role in mathematics, engineer and physics. This mathematical transformation takes a complex vector of length $N$, $\left(x_0, x_1, ..., x_{N-1}\right)$ and transforms it into another complex vector of the same length, $\left(y_0, y_1,...y_{n-1}\right)$ whose $k-th$ element is defined as
\begin{equation}
y_k\equiv \frac{1}{\sqrt{N}}\sum_{j=0}^{N-1}x_j e^{2\pi i j}.
\end{equation}

\begin{figure}
\centering
\includegraphics[width=0.48\textwidth]{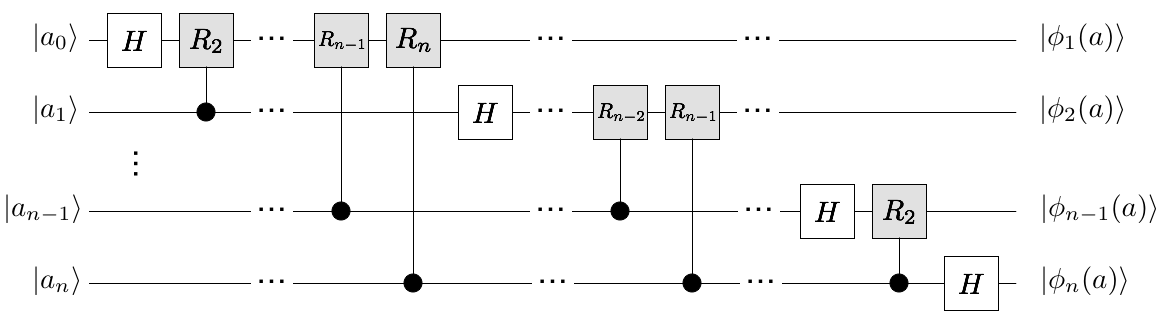} 
\caption{{\bf Digital implementation of the Q$\mathcal{F}$T for an $n-$~qubit system}. The single-qubit gate $H$ corresponds to the Hadamard gate (see Eq.~\ref{eq: H}). The rest are the controlled rotations defined by $cR_k=\dyad{0}\otimes\mathbbm{1}+\dyad{1}\otimes R_k$, where $R_k=\dyad{0}+e^{2\pi i/2^k}\dyad{1}$. The swap gates at the end of the circuit needed to correctly read the transformed state are not shown.}\label{fig:QFT}
\end{figure}

Q$\mathcal{F}$T, its quantum counterpart, is a linear operator, $\mathcal{F}$, with the following action on the basis states
\begin{equation}
\mathcal{F}\ket{	\Omega}\equiv \frac{1}{\sqrt{N}}\sum_{k=0}^{N-1} e^{2\pi i \Omega k/N}\ket{k},
\end{equation}
where $N=2^n$ and $n$ is the number of qubits of the system. The quantum-circuit implementation of the Q$\mathcal{F}$T is depicted in Fig.~\ref{fig:QFT}. The only single-qubit gates applied are Hadamard gates $H$, whose unitary matrix and Hamiltonian expressions are
\begin{equation}\label{eq: H}
H=e^{iH_H}=\frac{1}{\sqrt{2}}\begin{pmatrix}
1&1\\
1&-1
\end{pmatrix},~~H_H=\frac{\pi}{2}\left[\mathbbm{1}-\frac{1}{\sqrt{2}}(Z+X)\right],
\end{equation}
respectively. The entangling two-qubit gates of the circuit implementation are the controlled-$R_k$ operations, with
\begin{equation}
R_k=\begin{pmatrix}
1&0\\
0&e^{2\pi i/2^k}
\end{pmatrix},
\end{equation}
\begin{equation}\label{eq: cR_k}
cR_k=\dyad{0}\otimes\mathbbm{1}+\dyad{1}\otimes R_k={\small\begin{pmatrix}
1&0&0&0\\
0&1&0&0\\
0&0&1&0\\
0&0&0&e^{2\pi i/2^k}
\end{pmatrix}}.
\end{equation}
They appear in $(n-1)$ different blocks of controlled rotations, all of them preceded by a Hadamard gate, as shown in Fig.~\ref{fig:QFT}.

In order to apply the DAQC protocol to implement the Q$\mathcal{F}$T, we express the unitary matrices defined in Eq.~\eqref{eq: cR_k} in terms of an inhomogeneous ATA 2-body Ising Hamiltonian. Indeed,
\begin{equation} \label{eq: U_QFT}
U_{\text{Q$\mathcal{F}$T}}=\left[\prod_{m=1}^{n-1}U_{\text{SQG},m}U_{\text{TQG},m}\right]\cdot U_{H,m},
\end{equation}
where
\begin{eqnarray}
U_{\text{SQG},m}&=&\exp\left[i\sum_{k=2}^{n-(m-1)} \theta_k\left(\mathbbm{1}_{N\times N}-Z^{(k+m-1)}-Z^{(m)}\right)\right]\times\nonumber\\
&&\times \exp\left[ \frac{i\pi}{2}\left(\mathbbm{1}-\frac{Z^{(m)}+X^{(m)}}{\sqrt{2}}\right)\right], \label{eq:U_SQG}\\
U_{\text{TQG}}&=&\exp\left(i\sum_{c<k}^n\alpha_{c,k,m}Z^{(c)}\otimes Z^{(k)}\right),\label{eq:H_ZZ}\\
U_{H,m}&=&\exp\left(\frac{i \pi}{2}\left[\mathbbm{1}^{(m)}-\frac{(Z^{(m)}+X^{(m)})}{\sqrt{2}}\right]\right),\label{eq:U_Hm}\\
\theta_k&=&\frac{\pi}{2^{k+1}}\qquad\text{and}\qquad\alpha_{c,k,m}=\delta_{c,m}\frac{\pi}{2^{k-m+2}}.\label{eq:theta_alpha}
\end{eqnarray}
The superindices in brackets specify the qubit in which the unitary operation is performed. 

\begin{figure*}
\centering
\includegraphics[width=0.9\textwidth]{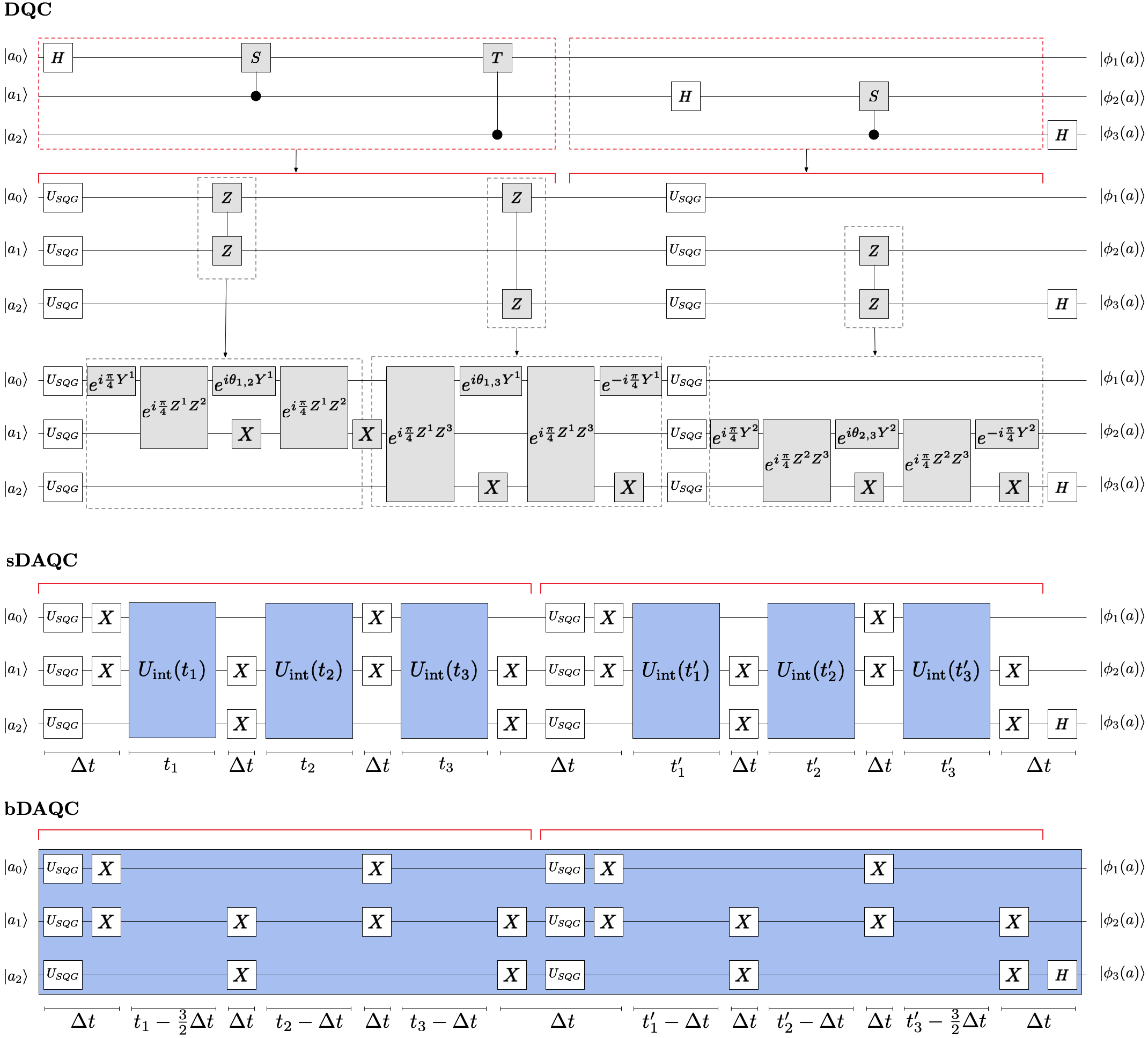} 
\caption{{\bf Implementation of the Q$\mathcal{F}$T for a $3-$~qubit system using three different protocols: DQC, sDAQC and bDAQC}. {\bf Digital implementation}: We show the transformation between the usual DQC implementation of the Q$\mathcal{F}$T (see Fig.~\ref{fig:QFT}) and the one that follows the Hamiltonian described by Eq.~\eqref{eq:H_ZZ}. Following Eq.~\eqref{eq: cR_k}, the controlled-rotation $cR_2$ and $cR_3$ correspond to the controlled-phase gate $cS$, and the controlled-$\pi/8$ gate $cT$, respectively. For the implementation that follows the Hamiltonian described by Eq.~\eqref{eq:H_ZZ}, each entangling two-qubit gate is applied according to the ATA DQC protocol, using a fixed $\pi/4$ phase (see Eq.~\ref{eq:ZZ gates}). {\bf DAQC Implementation}: The blue blocks $U_\text{int}(t)$ represent the analog blocks and each of them is applied during different times, $t$. The single-qubit gates $X$ refer to the Pauli matrix $\sigma_x$ and act for a time $\Delta t$. We apply the DAQC protocol for each block of controlled rotations of the DQC implementation, which is detailed by the red line over each of those blocks. The sDAQC switches on and off the analog evolution before applying the single-qubit rotations $X$. In contrast, in the bDAQC protocol, the single-qubit rotations are performed on top of the analog evolution. Since we are applying a Suzuki-Lie-Trotter decomposition to minimize the error, between the single-qubit rotations, each analog block acts for different times $t_i-\Delta t$, except for the first and the last block, which act for times $t_i-\frac{3}{2}\Delta t$.} \label{fig:DQC_sDAQC_bDAQC}
\end{figure*}

In Fig.~\ref{fig:DQC_sDAQC_bDAQC}, we depict the DQC implementation of the Q$\mathcal{F}$T using Eqs.~(\ref{eq:U_SQG}-\ref{eq:U_Hm}). As one can see, each controlled-rotation block can be implemented by applying first a set of single-qubit gates, and then a set of two-qubit gates. This is why we decompose the complete unitary transformation into three different operations. The subindices SQG and TQG stand for single qubit gates and two qubit gates, respectively.

Every two-qubit gate is applied following the ATA DQC protocol and using a fixed $\pi/4$ phase
\begin{equation}
e^{i\varphi_{jk}^{\mu\nu}\sigma_\mu^j\sigma_\nu^k}=e^{i\frac{\pi}{4}\sigma_y^j}e^{i\frac{\pi}{4}\sigma_\mu^j\sigma_\nu^k}e^{i\varphi_{jk}^{\mu\nu}\sigma_y^j}e^{-i\frac{\pi}{4}\sigma_\mu^j\sigma_\nu^k} e^{-i\frac{\pi}{4}\sigma_y^j}.
\end{equation}
In our case, $\mu=\nu=Z$ and the phase $\varphi_{jk}^{\mu\nu}$ correspond to the coefficient $\alpha_{c,k,m}$, given in Eq.~\eqref{eq:theta_alpha},
\begin{equation} \label{eq:ZZ gates}
e^{i\alpha_{c,k,m}Z ^c Z ^k}=e^{i\frac{\pi}{4}Y^c}e^{i\frac{\pi}{4}Z^c Z^k}e^{i\alpha_{c,k,m} Y^c} X^k e^{i\frac{\pi}{4} Z^c Z^k} X^k e^{-i\frac{\pi}{4} Y^c}.
\end{equation}

The inhomogeneous ATA 2-body Ising Hamiltonian which we want to write in the DAQC framework (see Eq.~\ref{eq:H_ZZ}) represents a complete block of controlled-rotations and it is different for each block. This means that we need to apply the DAQC protocol $(n-1)$ times, one time per controlled-rotation block, as depicted in Fig.~\ref{fig:DQC_sDAQC_bDAQC}.

\begin{figure*}
\centering
\includegraphics[width=\textwidth]{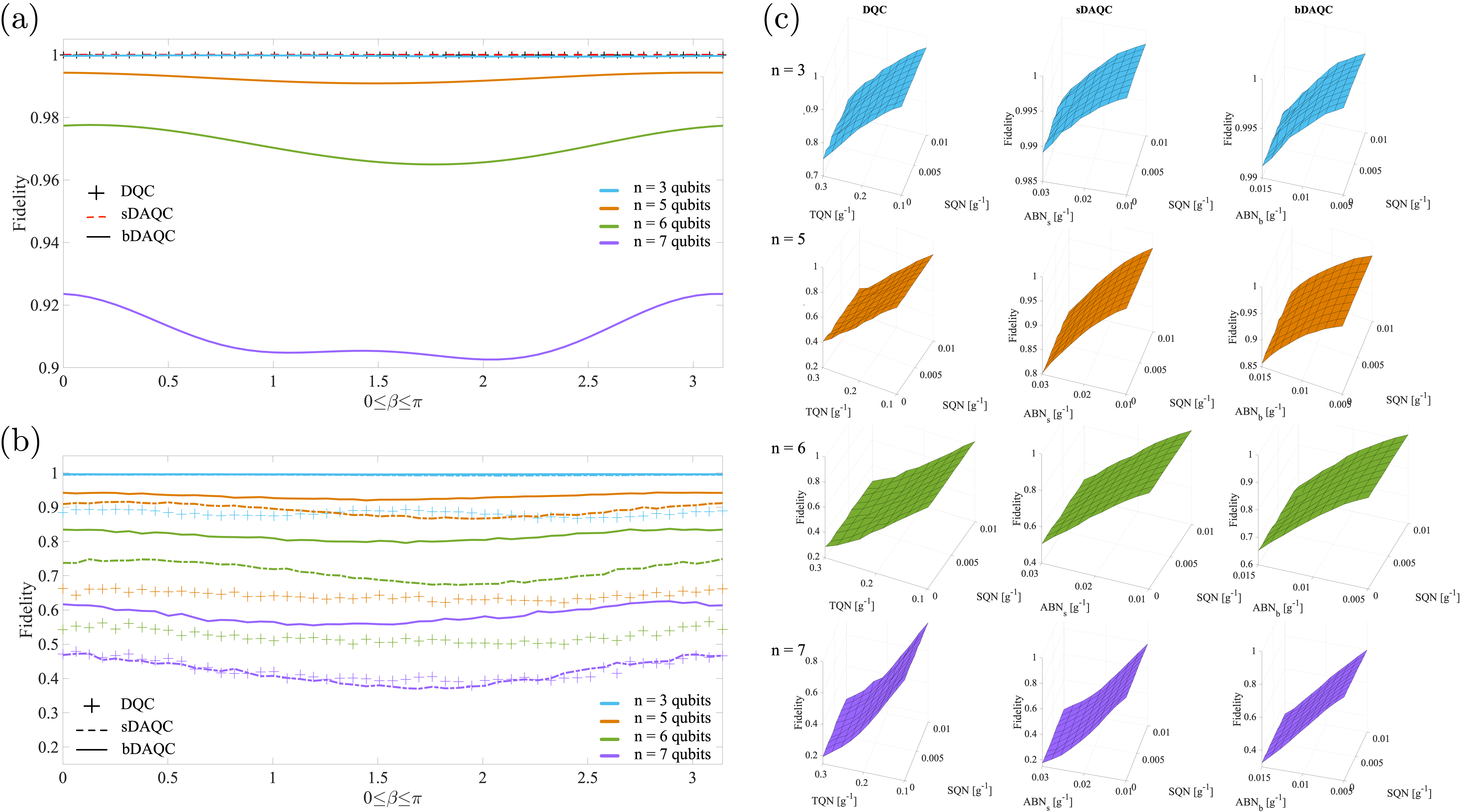} 
\caption{{\bf Fidelity of the transformation of the family of states $\ket{\psi_0}=\sin\beta\ket{W_n}+\cos\beta\ket{GHZ_n}$ using the three protocols. (a) Ideal implementation}. Both the DQC and the sDAQC protocols perform the Q$\mathcal{F}$T with fidelity $F_{\text{DQC}}=F_{\text{sDAQC}}=1$. The bDAQC has an intrinsic error due to the fact that the analog block is applied during the whole process. In this case, the fidelity is $F_{\text{bDAQC}}\in(0.90,1)$. {\bf (b) Realistic noisy implementation}. In this situation, the intrinsic error of the bDAQC is less significant than the experimental errors of the DQC and the sDAQC. The fidelity of the DQC decreases fast with the number of qubits. For a $6-$qubit system, the fidelity is around the $50\%$, so the DQC protocol is no longer useful. The fidelity of both DAQC protocols behaves better than the one of the DQC with the number of qubits, and remains above $70\%$ for the sDAQC and over $80\%$ for the bDAQC. For the case of a $7-$qubit system, we obtain a similar fidelity for both the DQC and the sDAQC protocol, but the one obtained for the bDAQC remains remarkably higher. This shows that the bDAQC protocol is the best option if one wants to implement the Q$\mathcal{F}$T on a system built up from several qubits.{\bf (c) Fidelity evolution with growing errors}. We want to show how the fidelity behaves if the errors we have estimated are slightly different. We have computed the $Q\mathcal{F}$T of the state $\ket{\psi_0}=\sin\frac{\pi}{4}\ket{W_n}+\cos\frac{\pi}{4}\ket{GHZ_n}$ for a system of $3$, $5$, $6$ and $7$ qubits. The fidelity of the two DAQC protocols is better than the fidelity obtained with the DQC, no matter what the errors are. Note that, in the DQC protocol, the error of the two-qubit gates dominates in the total error. Similarly, the fidelity in the DAQC is mainly affected by the errors in the analog blocks.}
\label{fig:Fidelity_vs_growing errors}\label{fig:beta states}
\end{figure*}

In order to compare each protocol (DQC, sDAQC and bDAQC), we compute the Q$\mathcal{F}$T of the family of states $\ket{\psi_0}=\sin\beta\ket{W_n}+\cos\beta\ket{GHZ_n}$, where $\beta$ runs from $0$ to $\pi$ and $n$ refers to the number of qubits of the system. We perform this for a $3-$, $5-$, $6$ and $7-$qubit system to grasp the behavior of the fidelity when the number of qubits scales up. As a figure of merit, we have calculated the fidelity between the states after the exact transformation and the ones obtained by the applied different methods,
\begin{equation}
F_{\text{method}}=\left|\braket{\psi_\mathcal{F}^{\text{exact}}}{\psi_\mathcal{F}^{\text{method}}}\right|^2.
\end{equation}
The results obtained are depicted in Fig.~\ref{fig:beta states}(a). According to the aforementioned arguments, the expected fidelity for both the digital case and the stepwise case is $F_{\text{DQC}}=F_{\text{sDAQC}}=1$, since the implementation is exact and ideal. This holds independently of the number of qubits of the system. The fidelity obtained when applying the bDAQC is always $F_{\text{bDAQC}}<1$, due to the intrinsic error associated to this method. The fidelity decreases with the number of qubits, but $F_{\text{bDAQC}}^{\text{3, 5, 6, 7}}>0.90$ for $n=3, 5, 6, 7$ qubits.

\section{Realistic implementation with experimental errors}\label{sec:noise}
Impurities in the materials comprising superconductive circuits and spurious interactions among superconducting qubits (cross-talk) and with two-level fluctuators modify the dynamics of the system, directly affecting the results of an experiment. Additionally, there are relevant control errors in the pulses when applying the gates. In order to make a fair comparison among different methods, we must introduce the effects of errors in the dynamics.

In single-qubit gates, we have introduced a magnetic field noise $\Delta B_\gamma$ by adding to the Hamiltonian of the single-qubit gate a random variable taken from a uniform probability distribution centered in $1$, \ie $\mathcal{U}(1-\text{SQGN}, 1+\text{SQGN})$. We have chosen SQGN~$=0.0005$. For the two-qubit gates, we add a Gaussian phase noise $\epsilon\in\mathcal{N}(0,\text{TQGN})$, with variance TQGN~$=0.2000$, to the $\pi/4$ phases in the DQC protocol. Finally, to model the experimental control error on the analog blocks, we include a Gaussian coherent noise to the time those blocks are applied, this is $t\rightarrow t+\delta$, where $\delta\in\mathcal{N}(0,\text{ABN})$. The value of the variance ABN depends on which DAQC protocol we are using. The value used on the sDAQC is double the value used for the bDAQC case. The values we have considered are $\text{ABN}_\text{s}=0.0200$ for the sDAQC case and $\text{ABN}_\text{b}=0.0100$ for the bDAQC case. Thus, each ideal gate transforms as
\begin{eqnarray}
e^{i\theta_k Z}&\rightarrow& e^{i\theta_k \Delta B\cdot Z},\\
e^{i\frac{\pi}{4}Z Z} &\rightarrow&e^{i\frac{\pi}{4}(1+\epsilon)Z Z},\\
e^{it_\alpha H_{int}}&\rightarrow&e^{i(t_\alpha +\delta) H_{\text{int}}}.
\end{eqnarray}
To test how the fidelity behaves for each case, we have computed the Q$\mathcal{F}$T of the family of states $\ket{\psi_0}=\sin\beta\ket{W_n}+\cos\beta\ket{GHZ_n}$. We repeated the simulation $1000$ times and calculate the average for a $3-$, $5-$, $6$ and $7-$qubit system. Both the sDAQC and the bDAQC perform better than the DQC protocol under realistic conditions, as depicted in Fig.~\ref{fig:beta states}(b). The best result corresponds to the bDAQC, a completely different situation to the ideal case. The reason is that, turning on and off the interaction Hamiltonian, is much more prone to suffer from experimental errors than keeping it on until the end of the computation. Note that the fidelity of the DQC decreases faster than the fidelity of the DAQC protocols with the number of qubits of the system. An extrapolation of these results to larger systems might indicate the convenience of this paradigm to foster near-term quantum computation. This shows the convenience of this paradigm to foster near-term quantum computation. The reason why the fidelity in Fig.~\ref{fig:beta states}(b) decays so fast with the number of qubits $N$, is the quadratic increment $\mathcal{O}(N^2)$ in the depth of the algorithm. This is an exponential speed-up with respect to the classical algorithm, but the errors accumulate very fast with the system size. As a consequence, the implementation of the Q$\mathcal{F}$T, which is a key subroutine in many other quantum algorithms, is a challenge when compared against algorithms with logarithmic scaling. This shows the importance of exploring alternative approaches, such as our DAQC, to optimize the implementation of this quantum algorithm.

Additionally we have studied how the fidelity behaves with different values for the errors. We have computed the Q$\mathcal{F}$T of the state $\ket{\psi_0}=\sin\frac{\pi}{4}\ket{W_n}+\cos\frac{\pi}{4}\ket{\text{GHZ}_n}$ for $n=3, 5, 6\text{ and }7$ qubits employing the three protocols. Again, the fidelity obtained when we use the bDAQC protocol behaves better than the one obtained by using the DQC protocol. For the $3-$qubit case, the fidelity is higher than $99\%$, for $5$ qubits it stays above $85\%$, and for $6$ qubits, it is greater than $70\%$. For $7$ qubits we obtain significantly lower fidelity than in the previous systems for the three protocols, but the one offered by the bDAQC remains above those offered by the DQC and the sDAQC. These results are depicted in Fig.~\ref{fig:Fidelity_vs_growing errors}(c). Similarly to the DQC case, in which the error of the two-qubit gates dominates in the total error, the fidelity in the DAQC is mainly affected by the errors in the analog blocks. The results of the simulations shown in Fig.~\ref{fig:Fidelity_vs_growing errors} should not be understood as the expected fidelity arising from an experiment, since it will strongly depend on the fabrication, architecture, and materials, among other factors. The actual result is that the distance between the fidelities of the bDAQC and the DQC are always remarkable in favor of the bDAQC. Even though we have considered a sensible choice for the noise model of the DAQC implementation, this model will not be accurate until comparing it against experimental data.

We would like to point out that, although the connectivity of most quantum systems is not ATA, with the remarkable exception of trapped ions and NMR, we have chosen it for the sake of simplicity. It can be shown that, using the DAQC paradigm, one can simulate a $N$-qubit ATA Ising Hamiltonian making use of, at most, $\frac{1}{2}N(N-1)$ nearest-neighbor (NN) Hamiltonians. This quadratic overhead holds in the worst case scenario, but it can be substantially reduced if there exists some pattern in our couplings (see Appendix). However, this overhead (or worse) also holds for the DQC paradigm. Consequently, the values of the fidelities could change in case of considering NN Hamiltonians, but the comparison shown between DQC and DAQC is fair and the conclusion related to the better performance of the bDAQC with respect to the DQC still holds.

\section{Conclusions and perspectives}
We have shown that the DAQC paradigm can enhance the depth of the implementation of quantum algorithm. In particular, we have provided a digital-analog algorithm for the Q$\mathcal{F}$T, an ubiquitous quantum subroutine, which is a relevant part of several quantum algorithms. Improving the fidelity of the implementation of Q$\mathcal{F}$T consequently enhances applicability of other quantum algorithms, such as Shor's algorithm for prime number factorization or HHL algorithm for solving linear systems of equations.

The main problem of the digital approach for the Q$\mathcal{F}$T in a real NISQ chip is that its fidelity decays fast when scaling up, since the depth of the algorithm grows quadratically with the number of qubits. Although there exist QEM techniques to reduce the DQC implementations, they are restricted to short-depth quantum algorithms, which constrains the achievable quantum volume of the algorithms, consequently restricting the original problem to a size in which DQC offers a reliable fidelity. In this manuscript, we have shown that DAQC could allow us to attain larger algorithm volumes, while keeping the fidelity under control. Indeed, we have simulated the Q$\mathcal{F}$T for $3$, $5$, $6$, and $7$ qubits, keeping the fidelity of the algorithm above $80\%$. In a similar situation, the fidelity provided by the DQC protocol is between $50\%$ and $65\%$. 

As a future work, it would be useful to include other types of errors, such as decoherence, and study the behavior of the fidelity with these errors. Taking into account the advantages that the DAQC paradigm offers, the next step is to study the implementation of quantum algorithms comprising the Q$\mathcal{F}$T as a subroutine. Additionally, it would also be interesting to implement other quantum algorithms such as Grover's algorithm. The successful implementation of these algorithms would pave the way for achieving useful quantum supremacy in the NISQ era.

The authors acknowledge support from Spanish MCIU/AEI/FEDER (PGC2018-095113-B-I00), Basque Government IT986-16, the projects QMiCS (820505) and OpenSuperQ (820363) of the EU Flagship on Quantum Technologies and the EU FET Open Grant Quromorphic. This work is supported by the U.S. Department of Energy, Office of Science, Office of Advanced Scientific Computing Research (ASCR) quantum algorithm teams program, under field work proposal number ERKJ333. 

\appendix*
\section{\uppercase{Enhancing connectivity in quantum processors}}\label{App.}
The connectivity of many quantum systems is not ATA, as assumed in this manuscript. For the aim of this work, this is not important, since we are trying to compare the performance of the DAQC approach against the DQC one, and we are also assuming ATA gates for the DQC. The choice of the ATA connectivity in this manuscript is therefore made for the sake of simplicity, since it can be shown \cite{ASIERTFG} that, using the DAQC paradigm, one can simulate a $L$-qubit ATA Ising Hamiltonian making use of, at most, $L(L-1)/2$ nearest-neighbor (NN) Hamiltonians. Indeed, the quadratic overhead holds in the worst case scenario, but it can be substantially reduced if there exists any pattern in the couplings \cite{ASIERTFG}. Consequently, this transformation from NN to ATA layout is physically equivalent to enhance the connectivity of the chip in the software level, without changing the architecture of the hardware or employing ancillary qubits. In the following, we present a brief description on how to proceed to attain this transformation. For a more detailed explanation, see Ref. \cite{ASIERTFG}.

We can understand the Ising Hamiltonian for $L$ qubits as a weighted graph of $L$ vertices, where the weight of the edge that connect the vertex $i$ to the vertex $j$ is the coupling constant $g_{ij}$. If two vertices $i$ and $j$ are not connected, then $g_{ij}=0$. Using this interpretation, an ATA Ising Hamiltonian of $L$ qubits corresponds to a complete graph $K_L$, which is a graph with edges among every possible vertex, with no repetition. The NN Ising Hamiltonian is then represented as a path that visits all the possible vertices only onces. This is called {\it Hamiltonian path} (HP). The problem reduces to the decomposition of a $K_L$ graph into a set of HP, using a NN Hamiltonian as a resource. For simplicity, we will assume that we are dealing with unweighted graphs, which are equivalent to homogeneous Ising Hamiltonians.

To obtain an efficient decomposition, we require a set of at most $L(L-1)/2$ HPs. This set of $k$ HPs is generated by a vertex permutation $P$, given by
\begin{equation}
P_L^k(j) =     \left\{ \begin{array}{rcl} (k-1+\frac{j}{2} \text{ mod L+1}, & \text{if } j \text{ even},\\
\\
															(k-1-\frac{j-1}{2} \text{ mod L+1}, & \text{if } j \text{ odd},
    	                \end{array}\right. 
\end{equation}
with $L$ the number of qubits, $k\in\mathbb{Z}$ such that $0\leq k\leq L/2$ and $j$ represents the $j$-th position of the vertex permutation.

We will obtain each HP employing a NN Hamiltonian as a resource. For this, we will change the connection of the resource using a iSWAP gate, which performs the following operations:
\begin{eqnarray}
U(Z\otimes \mathbbm{1})U^\dagger&=&\mathbbm{1}\otimes Z,\nonumber\\
U(\mathbbm{1}\otimes Z)U^\dagger&=&Z\otimes \mathbbm{1}.
\end{eqnarray}
The iSWAP gate, $U_{ij}$, changes a gate $Z$ acting on qubit $i$ to act on qubit $j$. Thus, if we sandwich a $Z^{(k)}Z^{(l)}$ term with the $U_{ij}$ iSWAP gate, we obtain
\begin{equation}
U_{ij}Z^{(k)}Z^{(l)}U_{ij}^\dagger=Z^{(\tau_{ij}(k))}Z^{(\tau_{ij}(l))},
\end{equation}
where $\tau_{ij}$ represents the permutation of the indices $i$ and $j$. If $k\neq i,j$, then $\tau_{ij}(k)=k$, otherwise $\tau_{ij}(i)=j$ and $\tau_{ij}(j)=i$.

Without applying the iSWAP gate, the initial vertex permutation of the system is $P=[1, 2, 3, \ldots, L]$. After the iSWAP operation, the system is defined by the new permutation $P'=[\tau_{ij}(1), \tau_{ij}(2), \tau_{ij}(3), \ldots, \tau_{ij}(L)]$. This approach can be generalized to a system with arbitrary connections. 

To sum up, the ATA Ising Hamiltonian $K_L$ graph is the sum of, at most, $L(L-1)/2$ HP, each of them built by applying different permutations to a NN Ising Hamiltonian graph.


\begin{thebibliography}{X}
\bibitem{Manin1980}
Y. I. Manin, {\it Vychislimoe i nevychislimoe} (Sov. radio, 1980).

\bibitem{Feyn1982}
R. P. Feynman, {\it Simulating physics with computers}. International journal of theoretical physics {\bf 21}, 467 (1982).

\bibitem{Shor1996}
P. W. Shor, {\it Polynomial-Time Algorithms for Prime Factorization and Discrete Logarithms on a Quantum Computer}. SIAM Review {\bf 26}, 1484 (1996).

\bibitem{Grover1996}
L. K. Grover, {\it Proceedings of the 28th Annual ACM Symposium on the Theory of Computing (STOC)}. ACM, New York (1996). 

\bibitem{QAlgsBeginers2018}
P. J. Coles, S. Eidenbenz, S. Pakin, A. Adedoyin, J. Ambrosiano, P. Anisimov, W. Casper, G. Chennupati, C. Coffrin, H. Djidjev, D. Gunter, S. Karra, N. Lemons, S. Lin, A. Lokhov, A. Malyzhenkov, D. Mascarenas, S. Mniszewski, B. Nadiga, D.~O'Malley, D. Oyen, L. Prasad, R. Roberts, P. Romero, N.~Santhi, N. Sinitsyn, P. Swart, M. Vuffray, J. Wendelberger, B. Yoon, R. Zamora, and W. Zhu, {\it Quantum algorithm implementations for beginners}. arXiv:1804.03719 (2018).

\bibitem{MCRMCLOSS2019} 
A. Martin, B. Candelas, \'A. Rodr\'iguez-Rozas, J. D. Mart\'in-Guerrero, X. Chen, L. Lamata, R. Or\'us, E. Solano, and M.~Sanz, {\it Towards Pricing Financial Derivatives with an IBM Quantum Computer}. arXiv:1904.05803 (2019).


\bibitem{TBG17}
K. Temme, S. Bravyi, and J.M. Gambetta, {\it Error Mitigation for Short-Depth Quantum Circuits}. Physical Review Letters {\bf 119}, 180509 (2017).

\bibitem{EBL18}
S. Endo, S. C. Benjamin, and Y. Li, {\it Practical Quantum Error Mitigation for Near-Future Applications}. Physical Review X {\bf 8}, 031027 (2018).

\bibitem{CTGSTGC19}
A. Chiesa, F. Tacchino, M. Grossi, P. Santini, I. Tavernelli, D. Gerace, and S. Carretta, {\it Quantum hardware simulating four-dimensional inelastic neutron scattering}. Nature Physics {\bf 15}, 455 (2019).

\bibitem{KTCMCG19}
A. Kandala, K. Temme, A. D. Co\' rcoles, A. Mezzacapo, J. M. Chow, and J. M. Gambetta, {\it Error mitigation extends the computational reach of a noisy quantum processor}. Nature {\bf 567}, 491 (2019).

\bibitem{MYB19}
S. McArdle, X. Yuan, and S. Benjamin, {\it Error-Mitigated Digital Quantum Simulation}. Physical Review Letters {\bf 122}, 180501 (2019).

\bibitem{PLLSS2018}
A. Parra-Rodriguez, P. Lougovski, L. Lamata, E. Solano, and M. Sanz, {\it Digital-Analog Quantum Computation}. arXiv:1812.03637 (2018).


\bibitem{DMNBT2002}
J. L. Dodd, Michael A. Nielsen, M. J. Bremner, and R. T. Thew, {\it Universal quantum computation and simulation using any entangling Hamiltonian and local unitaries}. Physical Review A {\bf 65}, 040301 (2002).

\bibitem{Nielsen2000} 
M. A. Nielsen and I. L. Chuang, {\it Quantum Computation and Quantum Information} (Cambridge University Press, Cambridge, 2000).

\bibitem{HHL2009}
A. W. Harrow, A. Hassidim, and S. Lloyd, {\it Quantum algorithm for linear systems of equations}. Physical Review Letters {\bf 103}, 150502 (2009).

\bibitem{LMR2014}
S. Lloyd, M. Mohseni, and P. Rebentrost, {\it Quantum Principal Component Analysis}. Nature Physics {\bf 10}, 631 (2014).

\bibitem{LHR2011}
B. P. Lanyon, C. Hempel, D. Nigg, M. Müller, R. Gerritsma, F.~Z\"ahringer, P. Schindler, J. T. Barreiro, M. Rambach, G.~ Kirchmair, M. Hennrich, P. Zoller, R. Blatt, and C. F. Roos, {\it Universal digital quantum simulation with trapped ions}. Science {\bf 334}, 57 (2011).

\bibitem{MMZB2016}
E. A. Martinez, C. A. Muschik, P. Schindler, D. Nigg, A. Erhard, M. Heyl, P. Hauke, M. Dalmonte, T. Monz, P. Zoller, and R. Blatt, {\it Real-time dynamics of lattice gauge theories with a few-qubit quantum computer}. Nature {\bf 534}, 516 (2016).
\bibitem{BLSM2015}
R. Barends, L. Lamata, J. Kelly, L. Garc\'ia-\'Alvarez, A.~G.~Fowler, A. Megrant, E. Jeffrey, T. C. White, D. Sank, J. Y. Mutus, B. Campbell, Y. Chen, Z. Chen, B. Chiaro, A.~Dunsworth, I.-C. Hoi, C. Neill, P. J. J. O'Malley, C. Quintana, P. Roushan, A. Vainsencher, J. Wenner, E. Solano, and J. M. Martinis, {\it Digital quantum simulation of fermionic models with a superconducting circuit}. Nature Communications {\bf 6}, 7654 (2015).

\bibitem{SMLS2015}
Y. Salath\'e, M. Mondal, M. Oppliger, J. Heinsoo, P. Kurpiers, A.~Potoc\v nik, A. Mezzacapo, U. Las Heras, L. Lamata, E.~Solano, S. Filipp, and A. Wallraff, {\it Digital quantum simulation of spin models with circuit quantum electrodynamics}. Physical Review X {\bf 5}, 021027 (2015).

\bibitem{BSLSNM2016}
R. Barends, A. Shabani, L. Lamata, J. Kelly, A. Mezzacapo, U.~Las Heras, R. Babbush, A. G.~Fowler, B. Campbell, Y.~Chen, Z. Chen, B. Chiaro, A. Dunsworth, E. Jeffrey, E. Lucero, A.~Megrant, J. Y. Mutus, M. Neeley, C. Neill, P. J. J. O'Malley, C. Quintana, P. Roushan, D. Sank, A. Vainsencher, J. Wenner, T. C. White, E. Solano, H. Neven, and J. M. Martinis, {\it Digitized adiabatic quantum computing with a superconducting circuit}. Nature {\bf 534}, 222 (2016).

\bibitem{LSKDBLTG2017}
N. K. Langford, R. Sagastizabal, M. Kounalakis, C. Dickel, A. Bruno, F. Luthi, D. J. Thoen, A. Endo, and L. DiCarlo, {\it Experimentally simulating the dynamics of quantum light and matter at deep-strong coupling}. Nature Communications {\bf 8} (2017).

\bibitem{KMTTBChG2017}
A. Kandala, A. Mezzacapo, K. Temme, M. Takita, M. Brink, J. M. Chow, and J. M. Gambetta, {\it Hardware-efficient variational quantum eigensolver for small molecules and quantum magnets}. Nature {\bf 549}, 242 (2017).

\bibitem{KDSLS2018}
N. Klco, E. F. Dumitrescu, A. J. McCaskey, T. D. Morris, R.~C.~Pooser, M. Sanz, E. Solano, P. Lougovski, and M.~J.~Savage, {\it Quantum-classical computation of Schwinger model dynamics using quantum computers}. Physical Review A {\bf 98}, 032331 (2018).

\bibitem{ARSL2018}
F. Albarr\' an-Arriagada, J. C. Retamal, E. Solano, and L. Lamata, {\it Measurement-based adaptation protocol with quantum reinforcement learning}. Physical Review A {\bf 98}, 042315 (2018).

\bibitem{OSCSL2018}
J. Olivares-S\' anchez, J. Casanova, E. Solano, and L.~Lamata, {\it Measurement-based adaptation protocol with quantum reinforcement learning in a Rigetti quantum computer}. arXiv:1811.07594 (2018).

\bibitem{DLMGLMS2019}
Y. Ding, L. Lamata, M. Sanz, J. D. Mart\' in-Guerrero, E. Lizaso, S. Mugel, X. Chen, R. Or\' us and E. Solano, {\it Towards Prediction of Financial Crashes with a D-Wave Quantum Computer}. arXiv:1904.05808 (2019).

\bibitem{SSSPS2016}
R. Sweke, M. Sanz, I. Sinayskiy, F. Petruccione, and E. Solano, {\it Digital quantum simulation of many-body non-Markovian dynamics}. Physical Review A {\bf 94}, 022317 (2016).

\bibitem{GALMSSL2016}
L. Garc\' ia-\' Alvarez, U. Las Heras, A. Mezzacapo, M. Sanz, E.~Solano, and L. Lamata, {\it Quantum chemistry and charge transport in biomolecules with superconducting circuits}. Scientific Reports {\bf 6}, 27836 (2016).

\bibitem{LSKDBLD2017}
N. K. Langford, R. Sagastizabal, M. Kounalakis, C. Dickel, A.~Bruno, F. Luthi, D. J. Thoen, A. Endo, and L. DiCarlo, {\it Experimentally simulating the dynamics of quantum light and matter at deep-strong coupling}. Nature Communications {\bf 8}, 1715 (2017).

\bibitem{BRGDS12}
D. Ballester, G. Romero, J. J. Garc\' ia-Ripoll, F. Deppe, and E. Solano, {\it Quantum Simulation of the Ultrastrong-Coupling Dynamics in Circuit Quantum Electrodynamics}. Physical Review {\bf X} 2, 021007 (2012).

\bibitem{MLHPDSL}
A. Mezzacapo, U. Las Heras, J. S. Pedernales, L. DiCarlo, E.~Solano, and L. Lamata, {\it Digital quantum Rabi and Dicke models in superconducting circuits}. Scientific reports {\bf 4}, 7482 (2014).

\bibitem{PLFRLS2015}
J. S. Pedernales, I. Lizuain, S. Felicetti, G. Romero, L. Lamata, and E. Solano, {\it Quantum Rabi model with trapped ions}. Scientific Reports {\bf 5}, 15472 (2015).

\bibitem{BMSSRWU17}
J. Braum\"uller, M. Marthaler, A. Schneider, A. Stehli, H. Rotzinger, M. Weides, and A. V. Ustinov, {\it Analog quantum simulation of the Rabi model in the ultra-strong coupling regime}. Nature Communications {\bf 8}, 779 (2017).

\bibitem{LALZPLSK2018}
D. Lv, S. An, Z. Liu, J.-N. Zhang, J. S. Pedernales, L. Lamata, E. Solano, and K. Kim, {\it Quantum Simulation of the Quantum Rabi Model in a Trapped Ion}. Physical Review X {\bf 8}, 021027 (2018).
\bibitem{MSLESS2015}
A. Mezzacapo, M. Sanz, L. Lamata, I. L. Egusquiza, S. Succi, and E. Solano, {\it Quantum simulator for transport phenomena in fluid flows}. Scientific Reports {\bf 5}, 13153 (2015).
\bibitem{FSLRJDS2014}
S. Felicetti, M. Sanz, L. Lamata, G. Romero, G. Johansson, P. Delsing, and E. Solano, {\it Dynamical Casimir effect entangles artificial atoms}. Physical Review Letters {\bf 113}, 093602 (2014).

\bibitem{RFERSS2016}
D. Z. Rossatto, S. Felicetti, H. Eneriz, E. Rico, M. Sanz, and E. Solano, {\it Entangling polaritons via dynamical Casimir effect in circuit quantum electrodynamics}. Physical Review B {\bf 93}, 094514 (2016).

\bibitem{SWGS2018}
M. Sanz, W. Wieczorek, S. Gröblacher, and E. Solano, {\it Electromechanical Casimir effect}. Quantum {\bf 2}, 91 (2018).
\bibitem{GACMELRS2015}
L. Garc\' ia-\' Alvarez, J. Casanova, A. Mezzacapo, I. L. Egusquiza, L. Lamata, G. Romero, and E. Solano, {\it Fermion-fermion scattering in quantum field theory with superconducting circuits}. Physical Review Letters {\bf 114}, 070502 (2015).

\bibitem{APLS2016}
I. Arrazola, J. S. Pedernales, L. Lamata, and E. Solano, {\it Digital-analog quantum simulation of spin models in trapped ions}. Scientific Reports {\bf 6}, 30534 (2016).

\bibitem{LPSS2018}
L. Lamata, A. Parra-Rodriguez, M. Sanz, and E. Solano, {\it Digital-Analog Quantum Simulations with Superconducting Circuits}. Advanced Physics X {\bf 3}, 1457981 (2018).

\bibitem{Deutsch1985}
D. Deutsch, {\it Quantum theory, the Church-Turing principle and the universal quantum computer}. Proceedings of the Royal Society of London A {\bf 400}, 97 (1985).

\bibitem{ASIERTFG}
A. Galicia, {\it Digital-Analog Quantum Computing}. Bachelor Thesis, UPV/EHU (2019).

\end{thebibliography}
\end{document}